\def \SAIT #1 #2 {{\em Mem.\ Soc.\ Astron.\ It.\/} {\bf #1}, #2}
\def \MESS #1 #2 {{\em The Messenger\/} {\bf #1}, #2}
\def \ASTRNACH #1 #2 {{\em Astron. Nach.\/} {\bf #1}, #2}
\def \AAP #1 #2 {{\em Astron. Astrophys.\/} {\bf #1}, #2}
\def \AAL #1 #2 {{\em Astron. Astrophys. Lett.\/} {\bf #1}, L#2}
\def \AAR #1 #2 {{\em Astron. Astrophys. Rev.\/} {\bf #1}, #2}
\def \AAS #1 #2 {{\em Astron. Astrophys. Suppl. Ser.\/} {\bf #1}, #2}
\def \AJ #1 #2 {{\em Astron. J.\/} {\bf #1}, #2}
\def \ANNREV #1 #2 {{\em Ann. Rev. Astron. Astrophys.\/} {\bf #1}, #2}
\def \APJ #1 #2 {{\em Astrophys. J.\/} {\bf #1}, #2}
\def \APJL #1 #2 {{\em Astrophys. J. Le tt.\/} {\bf #1}, L#2}
\def \APJS #1 #2 {{\em Astrophys. J. Suppl.\/} {\bf #1}, #2}
\def \APSS #1 #2 {{\em Astrophys. Space Sci.\/} {\bf #1}, #2}
\def \ASR #1 #2 {{\em Adv. Space Res.\/} {\bf #1}, #2}
\def \BAIC #1 #2 {{\em Bull. Astron. Inst. Czechosl.\/} {\bf #1}, #2}
\def \JSQRT #1 #2 {{\em J. Quant. Spectrosc. Radiat. Transfer\/} {\bf #1}, #2}
\def \MN #1 #2 {{\em Mon. Not. R. Astr. Soc.\/} {\bf #1}, #2}
\def \MEM #1 #2 {{\em Mem. R. Astr. Soc.\/} {\bf #1}, #2}
\def \PLR #1 #2 {{\em Phys. Lett. Rev.\/} {\bf #1}, #2}
\def \PASJ #1 #2 {{\em Publ. Astron. Soc. Japan\/} {\bf #1}, #2}
\def \PASP #1 #2 {{\em Publ. Astr. Soc. Pacific\/} {\bf #1}, #2}
\def \NAT #1 #2 {{\em Nature\/} {\bf #1}, #2}

\documentstyle{memsait}
\input epsf.sty
\begin{opening}
\title{ANTLIA: AN OUTSKIRT LOCAL GROUP GALAXY} 
\author{Marco Castellani$^1$, 
Giuseppe Bono$^1$, 
Santi Cassisi$^2$,
Gianni Marconi$^1$,
Anna Piersimoni$^2$,}
\institute{$^1$Osservatorio Astronomico di Roma, Italy\\
	   $^2$Osservatorio Astronomico di Collurania, Teramo, Italy\\	
}

\date{} 
\end{opening}

\begin{document}

\oddpagefooter{}{}{} 
\evenpagefooter{}{}{} 
\ 
\bigskip

\begin{abstract}

Deep (I,V-I) and (I,B-I) color-magnitude diagrams (CMDs) of the 
Antlia dwarf galaxy, based on Science Verification (SV) data collected with
the FORS I camera on the ESO Very Large Telescope (VLT) are presented. By
adopting the new calibration of the Tip of the Red Giant Branch (TRGB) provided
by Salaris \& Cassisi (1998) we estimated that the Antlia distance modulus is
(m-M)$_0$=25.98 $\pm$0.10 mag.
We suggest an improvement of the classical TRGB method based on infrared 
H and K magnitudes of TRGB stars. Such a method should overcome the
well-known limit of the "classical" TRGB method when dealing with 
metal-rich stellar populations.
\end{abstract}


\section{The "classical" Tip of the Red Giant Method}


\begin{figure}

\epsfxsize=12cm 
\hspace{3.5cm}\epsfbox{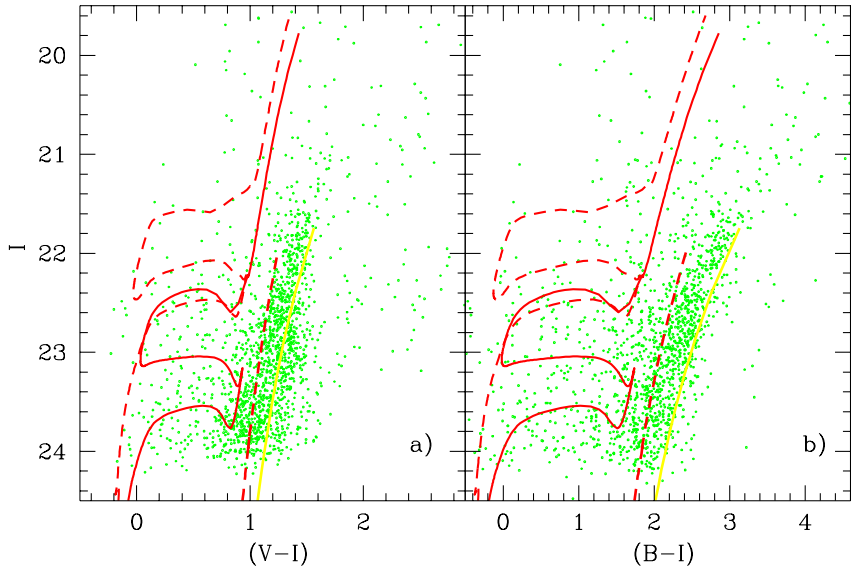} 
\vspace{-2cm}
\caption[h]{Antlia CM diagrams: I vs. (V-I) (left) and I vs. (B-I) (right)}
\end{figure}


Figure 1 shows the (I, V-I) and (I, B-I) CM diagrams of Antlia. 
As far as the data reduction is concerned, the interested reader is 
referred to Piersimoni et al. (1999). 
Data plotted in these CMDs present two key features: a well-defined 
Red Giant Branch, and a sample of bright blue stars, belonging to a young 
stellar component. By adopting the new calibration of the Tip of the
Red Giant Branch (TRGB) provided by Salaris \& Cassisi (1998) we estimated
that the Antlia distance modulus is (m-M)$_0$=25.98 $\pm$0.10 mag, 
equivalent to D=1.51$\pm$0.07 Mpc.

The main aim of this investigation is to present a new calibration 
of the TRGB method based on NIR magnitudes which can allow us to supply 
reliable distance determinations for both metal-poor {\em and} metal-rich
stellar populations. In fact empirical evidence suggest not only that the 
the mean metallicity of LG galaxies range from $[Fe/H]\approx=-2.2\pm0.1$ 
(Ursa Minor) to $[Fe/H]\approx=-0.8\pm0.1$ (NGC~205), but also that the average 
metallicity spread is of the order of 0.5 dex (Mateo 1998). Therefore,  
the TRGB method cannot be used to supply a homogeneous distance scale 
in the LG, since both theory and observations support the evidence that 
I magnitude of the TRGB stars is constant within 0.1 mag, for stellar 
populations with metal contents [Fe/H] ranging from -2.2 to -0.7 
(Lee, Freedman, \& Madore 1993; Salaris \& Cassisi 1997,1998).

In order to overcome this drawback and to improve the TRGB method in 
such a way that could be safely used to estimate the distances
of stellar systems which include both metal-poor and metal-rich 
stars, we investigated on theoretical grounds the properties of TRGB stars 
in the NIR bands. We focused our attention to NIR passbands because 
they present a negligible dependence on interstellar reddening and also 
a marginal dependence on metallicity.


\section{Discussion}

The behavior of TRGB magnitude in NIR bands (I,J,H,K) was investigated 
by adopting the theoretical predictions by  Salaris \& Cassisi (1997, 1998)
which were transformed into the observational plane using the bolometric 
corrections and the color-temperature relations provided by 
Wood \& Bessel (1994).

\begin{figure}
\epsfysize=12cm 
\hspace{3.5cm}\epsfbox{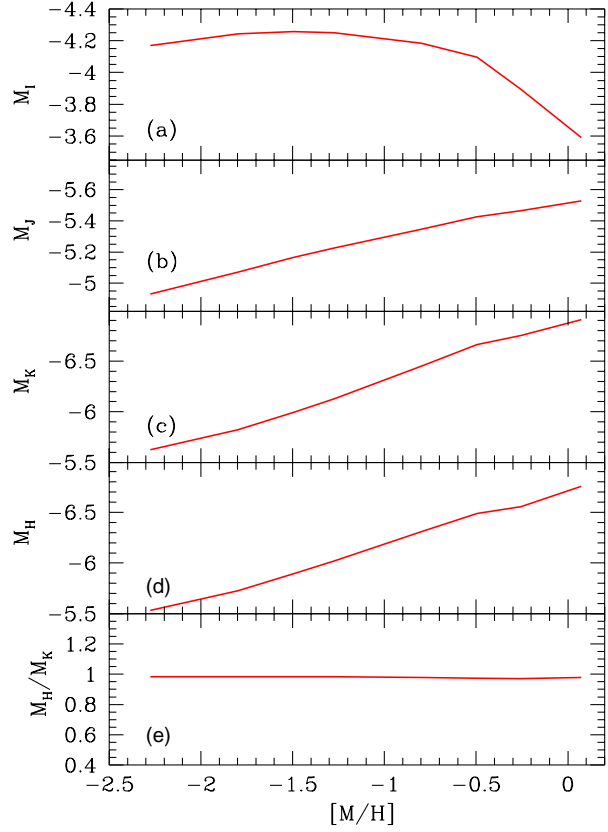} 
\caption[h]{TRGB magnitude as a function of global metallicity.}
\end{figure}


Panel (a) of Figure 2 shows the well known behavior of TRGB magnitudes in
the I band, i.e. the I magnitude of the tip is not a reliable standard candle 
for metallicities greater that [Fe/H]=-0.7. 
Panel (b) of the same figure shows the J band magnitude of TRGB stars. 
Data plotted in this panel clearly show that the TRGB magnitude in the 
J band cannot be adopted as a standard candle, since it shows a steady 
decrease when moving from metal-poor to metal rich structures.
The same outcome applies for H and K magnitudes (see panels c and d). 
A closer inspection, however, reveals that the behavior on this two
filters is quite similar and in particular that they present the same 
slope. This evidence prompted us to estimate the ratio M$_H$/M$_K$ 
as a function of metallicity.

Interestingly enough, we found that the ratio M$_H$/M$_K$ shows a 
maximum variance of approximately 0.01 mag over the entire metallicity 
range (see bottom panel) we have taken into account 
($-2.5 \le [M/H] \le 0.2$). This result suggests that HK ratio can 
be adopted as a standard candle, since it presents a negligible 
dependence on metallicity.


\section{A possible improvement of the TRGB method}

On the basis of the result of the previous section, we can assume that 
M$_H$/M$_K$ $\simeq$ const. = a.  Then, by accounting for the evidence 
that the extinction in the H and K bands are quite similar 
(for E$_{B-V}$ = 0.1, we have A$_H$-A$_K$$\simeq$0.02 mag), we obtain \\

\bigskip

\centerline{
m$_H$ - M$_H$ $\simeq$ m$_K$ - M$_K$. 
}

\bigskip

At this point, we can simply derive \\

\bigskip

\centerline{
M$_K$ $\simeq$ (m$_H$ - m$_K$) / ( a - 1) = ( H - K ) / ( a - 1 ).
}

\bigskip

This relation suggests that {\it dereddened measurements of the TRGB 
H and K magnitudes, allow us to estimate the distance modulus.} 
Once this "new" method is supported by empirical evidence it could  
be adopted to estimate distances of LG galaxies whose stellar 
populations are older few Gyrs.

Let us briefly mention some drawbacks of the distance indicators currently
adopted:

(1) The TRGB cannot be safely applied to all LG galaxies, mainly because of the
already quoted "metallicity problem".

(2) Due to the fact that the Horizontal Branch (HB) is intrinsically
several magnitude fainter than TRGB, distances based on $M_V$(HB) can be
derived only for nearby stellar systems.

(3) Distances based on "Cepheid PL relations" can be derived only for young 
stellar systems.

On the other hand, this new approach can be adopted 
in LG stellar systems characterized by
a wide range of ages and metallicities.
In fact, modern NIR detectors aboard on VLT (ISAAC) and on NTT (SOFI)
allow us to reach H and K magnitudes roughly equal to 19 mag.
By assuming a S/N=20, and a seeing of 1.0 arcsec, a limit magnitude K=19
can be reached within time exposures ranging from 650 sec (ISAAC) to 3000 sec (SOFI).
It turns out that the TRGB can be measured in
galaxies with distance moduli up to 
24-25. This means that this approach 
can be applied for approximately 70\% of southern
LG galaxies.

The "a" parameter of the previous equation has to be tested and calibrated 
on NIR data of stellar systems in the LG. Moreover, to test the intrinsic 
accuracy and reliability of this method we need to collect new NIR data  
for galactic globular clusters and LG dwarf galaxies whose distance is 
already well known from other distance indicators.








\end{document}